\documentclass[conference]{IEEEtran}
\IEEEoverridecommandlockouts
\usepackage{cite}
\usepackage{amsmath,amssymb,amsfonts}
\usepackage{algorithmic}
\usepackage{graphicx}
\usepackage{textcomp}
\usepackage{xcolor}
\def\BibTeX{{\rm B\kern-.05em{\sc i\kern-.025em b}\kern-.08em
    T\kern-.1667em\lower.7ex\hbox{E}\kern-.125emX}}

\usepackage{graphicx}
\usepackage{times}

\usepackage{amsmath,amsfonts,bm}

\def\eqref#1{equation~\ref{#1}}
\def\1{\bm{1}}

\DeclareMathAlphabet{\mathsfit}{\encodingdefault}{\sfdefault}{m}{sl}
\SetMathAlphabet{\mathsfit}{bold}{\encodingdefault}{\sfdefault}{bx}{n}

\usepackage{hyperref}
\usepackage{url}

\usepackage{comment}

\usepackage{tikz}
\usepackage{xcolor}
\usepackage{filecontents}
\usepackage{subcaption}
\usepackage{tablefootnote}
\usepackage{graphicx}
\usepackage{url}
\usepackage{hyperref}
\usepackage{comment}
\usepackage{color, soul}
\usepackage{mathrsfs}
\usepackage{cleveref}
\usepackage{multirow}
\usepackage{float}
\usepackage{wrapfig}
\usepackage{bm}
\usepackage{bbm}
\usepackage{algorithm}
\usepackage{algorithmic}
\usepackage{colortbl}
\usepackage{enumitem}
\usepackage{color}
\usepackage{balance}
\usepackage{stfloats}
\usepackage{makecell}
  \usepackage{pifont}

\usepackage{booktabs}       % professional-quality tables
\usepackage{nicefrac}       % compact symbols for 1/2, etc.
\usepackage{microtype}      % microtypography
\usepackage{xcolor}         % colors

\usepackage{microtype}
\usepackage{hyperref}
\usepackage{url}
\usepackage{booktabs}

\usepackage{lineno}

\definecolor{darkblue}{rgb}{0, 0, 0.5}
\hypersetup{colorlinks=true, citecolor=darkblue, linkcolor=darkblue, urlcolor=darkblue}

\begin{document}

\title{LSP-RAG: Latent Scoring Perspectives for Knowledge Graph Retrieval-Augmented Generation}

\author{
\IEEEauthorblockN{
Haochen Liu$^{1}$,
Guanghui Min$^{1}$,
Song Wang$^{2}$,
Yada Zhu$^{3}$,
Chen Chen$^{1}$,
Jundong Li$^{1}$
}
\IEEEauthorblockA{
$^{1}$University of Virginia \quad
$^{2}$University of Central Florida \quad
$^{3}$IBM Research\\
sat2pv@virginia.edu,
jjm8vr@virginia.edu,
song.wang@ucf.edu,\\
yzhu@us.ibm.com,
zrh6du@virginia.edu,
jundong@virginia.edu
}
}

\newcommand{\fix}{\marginpar{FIX}}
\newcommand{\new}{\marginpar{NEW}}
\maketitle

\begin{abstract}
Large Language Models (LLMs) have demonstrated strong capabilities in open-domain question answering, but often struggle with factual accuracy and multi-hop reasoning due to the incompleteness of the training corpora. A promising solution is Knowledge Graph Retrieval-Augmented Generation (KG-RAG), which supplements LLMs with structured knowledge retrieved from external knowledge graphs (KGs). However, existing KG-RAG methods either rely on large-scale language models (e.g., ChatGPT) to guide the retrieval process, which leads to high computational costs, or suffer from limited retrieval quality when using lightweight language models, particularly under multi-hop scenarios. We propose LSP-RAG (Latent Scoring Perspectives for Retrieval-Augmented Generation over Knowledge Graphs), a novel KG-RAG framework that enhances hop-wise KG knowledge retrieval through multi-perspective prompted scoring. Each latent scoring perspective is guided by a combination of two types of soft prompts: a \emph{perspective-specific soft prompt} induces a distinct prompt-conditioned scoring function over candidate triplets, while a \emph{contextual soft prompt} evolves with each reasoning hop by encoding the query and previously explored KG triplets, enabling the model to preserve consistency and relevance across multi-hop retrieval. Rather than assuming manually defined semantic roles, these latent perspectives are adaptively aggregated to capture heterogeneous relevance signals during retrieval. This design allows LSP-RAG to perform accurate and robust retrieval using lightweight language models. LSP-RAG achieves superior performance across multiple KG-based Question Answering benchmarks compared to existing retrieval systems, including those that rely on much larger language models, demonstrating its effectiveness under limited computational budgets.
\end{abstract}


\section{Introduction}

Large Language Models (LLMs)~\cite{gpt,llama} have demonstrated impressive capabilities across a wide range of natural language processing tasks, including text comprehension~\cite{lewis2020bart}, open-domain question answering~\cite{wei2022chain}, and natural language generation~\cite{cheng2023can}. Despite these advances, LLMs remain limited due to the incompleteness of knowledge encoded during pretraining, which restricts their ability to provide factually accurate and up-to-date answers~\cite{IKE,wang2023knowledge}. This limitation is particularly pronounced in knowledge-intensive scenarios that demand factual precision, reasoning over complex multi-hop relations, and adaptation to continually evolving information~\cite{IKE,wang2023knowledge}. Consequently, LLM generations often contain hallucinated content, outdated facts, or unverifiable claims, undermining their reliability in real-world applications.

To mitigate these issues, Retrieval-Augmented Generation (RAG)~\cite{gao2023retrieval} has been proposed to enhance LLM reliability by incorporating external knowledge during inference. RAG allows models to access up-to-date and factual information beyond what was memorized during pretraining. While early RAG systems primarily relied on unstructured text corpora, recent efforts have explored structured sources of knowledge for more precise and interpretable reasoning. Among these, Knowledge Graphs (KGs) offer a highly structured and semantically rich format, representing facts as \textit{triplets} in the format of $(\text{subject}, \text{relation}, \text{object})$. This format supports explicit relation modeling and multi-hop relational reasoning, making KGs especially suitable for question answering tasks that require factual consistency and transparency~\cite{sun2024think,jiang2024kg}. In the KG-RAG paradigm~\cite{sanmartin2024kg}, models typically follow a two-step process: (1) \textit{knowledge retrieval}, where a subgraph containing query-relevant triplets is constructed by expanding from entities appearing in the query; and (2) \textit{prompt integration}, where triplets in the retrieved subgraph are presented as input to the LLM as evidence for answer generation. The performance of KG-RAG methods largely relies on the ability to retrieve and rank relevant triplets effectively, especially under multi-hop settings where irrelevant information can accumulate and mislead reasoning.

\begin{figure}[t]

     \centering \includegraphics[width=0.9\linewidth]{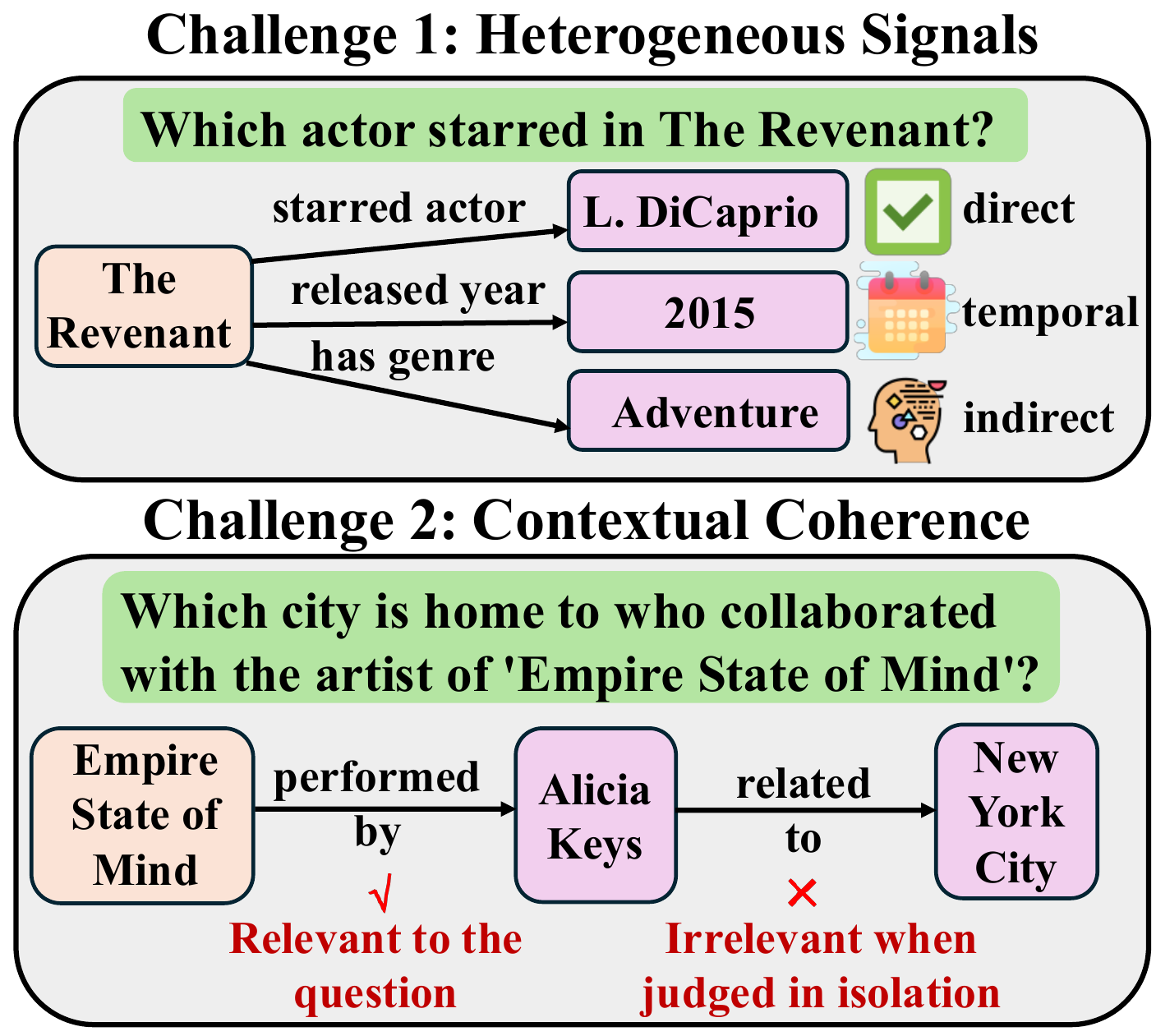}
    \caption{
Illustration of two key challenges in KG-RAG. In Challenge 1, candidate triplets from the same entity may reflect heterogeneous matching signals, such as direct factual relevance, temporal association, or indirect relational connection, making a single scoring strategy less flexible for effective retrieval.  
In Challenge 2, a relation like \texttt{related\ to} may appear irrelevant when judged in isolation, but becomes crucial when combined with prior steps. 
}
  \label{fig:intro}

\end{figure}

Despite recent progress, current KG-RAG approaches still face limitations that hinder their effectiveness. First, most methods assess candidate triplet relevance through a single lens, such as semantic similarity or prompt-based scoring~\cite{shi2023replug,sun2024think}. This rigid design fails to accommodate heterogeneous relevance signals across different candidate triplets, lacking the flexibility to account for diverse relational semantics~\cite{saxena2020improving}. A unified scoring strategy cannot effectively handle such diversity, leading to suboptimal triplet selection. Second, many candidate triplets may appear irrelevant when judged in isolation, especially those involving underspecified relations, but actually play a critical role in connecting earlier reasoning steps to the final answer. Existing methods typically treat each KG triplet independently and fail to model such multi-hop dependencies~\cite{feng2020scalable,zhang2022greaselm}. As a result, they struggle to maintain contextual continuity, often discarding useful evidence prematurely. Third, while most recent systems attempt to address these issues by leveraging large-scale language models such as ChatGPT to guide retrieval, which are computationally expensive and impractical for scalable or real-time applications~\cite{min2019discrete}. Examples of the first two limitations are shown in Figure~\ref{fig:intro}.

To address these challenges, we propose LSP-RAG, a lightweight but effective KG-RAG framework that enables lightweight language models to perform accurate and multi-hop retrieval. LSP-RAG contains a \textit{multi-perspective prompted scoring} module, which scores candidate KG triplets based on the query and previously explored knowledge. Each latent scoring perspective is parameterized by a learnable \emph{perspective-specific soft prompt}, inducing a distinct prompt-conditioned scoring function over candidate triplets. Since supervision is provided at the level of triplet relevance rather than manually defined perspective labels, these latent scoring perspectives are not intended to correspond to predefined semantic categories. Instead, they serve as complementary scoring components that are adaptively aggregated for flexible relevance estimation. In addition, LSP-RAG incorporates a \emph{contextual soft prompt} that is dynamically constructed from the triplets in previous reasoning steps. This prompt provides evolving contextual signals that help the scoring of new candidate triplets using earlier reasoning steps, allowing the model to assess each candidate not in isolation, but in light of the evidence already accumulated, addressing the limitation of prior work. By combining latent scoring perspectives with contextual adaptation, our model robustly filters irrelevant or noisy triplets. Notably, this framework is fully compatible with lightweight models, avoiding the reliance on expensive LLMs such as ChatGPT for retrieval, thereby offering both efficiency and strong performance. 
Our main contributions are listed as follows:

\begin{itemize}[leftmargin=0.15in]

\item We propose LSP-RAG, a KG-RAG framework based on latent scoring perspectives that enables lightweight language models to conduct multi-hop KG retrieval, reducing the cost of retrieval. 
\item We design two types of soft prompts to support our retrieval process: (1) \emph{perspective-specific soft prompt}, which induces multiple latent scoring perspectives over candidate triplets, and (2) \emph{contextual soft prompt} that incorporates evidence from the prior hop to support contextual retrieval. This combination brings both flexible relevance estimation and adaptivity into the retrieval process.
\item We conduct extensive experiments on KGQA benchmark datasets. LSP-RAG consistently outperforms existing KG-RAG baselines across all settings. Remarkably, it even surpasses ChatGPT retrieval despite using only lightweight language models for retrieval, achieving better accuracy and lower inference cost.
\end{itemize}

\section{Problem Formulation}

We study the task of Knowledge Graph Retrieval-Augmented Generation (KG-RAG), where structured knowledge from a knowledge graph is selectively retrieved to assist a language model in answering natural language questions. Formally, a knowledge graph is represented as $\mathcal{G} = (\mathcal{E}, \mathcal{R}, \mathcal{T})$, where $\mathcal{E}$ denotes the set of entities, $\mathcal{R}$ is the set of relations, and $\mathcal{T} = \{(h, r, t) \mid h, t \in \mathcal{E}, r \in \mathcal{R} \}$ is the set of directed triplets. Each triplet $(h,r,t)$ encodes the fact that head entity $h$ is connected to tail entity $t$ via relation $r$. Given a language model and a question $q$, the objective is to identify a subset of triplets from $\mathcal{T}$ that are relevant to answering $q$, and use them as evidence to enhance the language model's generation. Formally, if $\mathcal{T}_q^* \subseteq \mathcal{T}$ denotes the selected triplet set, the final answer $a_q$ is generated as:
\begin{equation}
a_q = \mathrm{LLM}\big(\,[q ; \mathcal{T}_q^*]\,\big).
\end{equation}
Here $\mathcal{T}_q^*$ contains the most relevant information for question answering.

\section{LSP-RAG}
\begin{figure*}[t]

     \centering \includegraphics[width=0.9\linewidth]{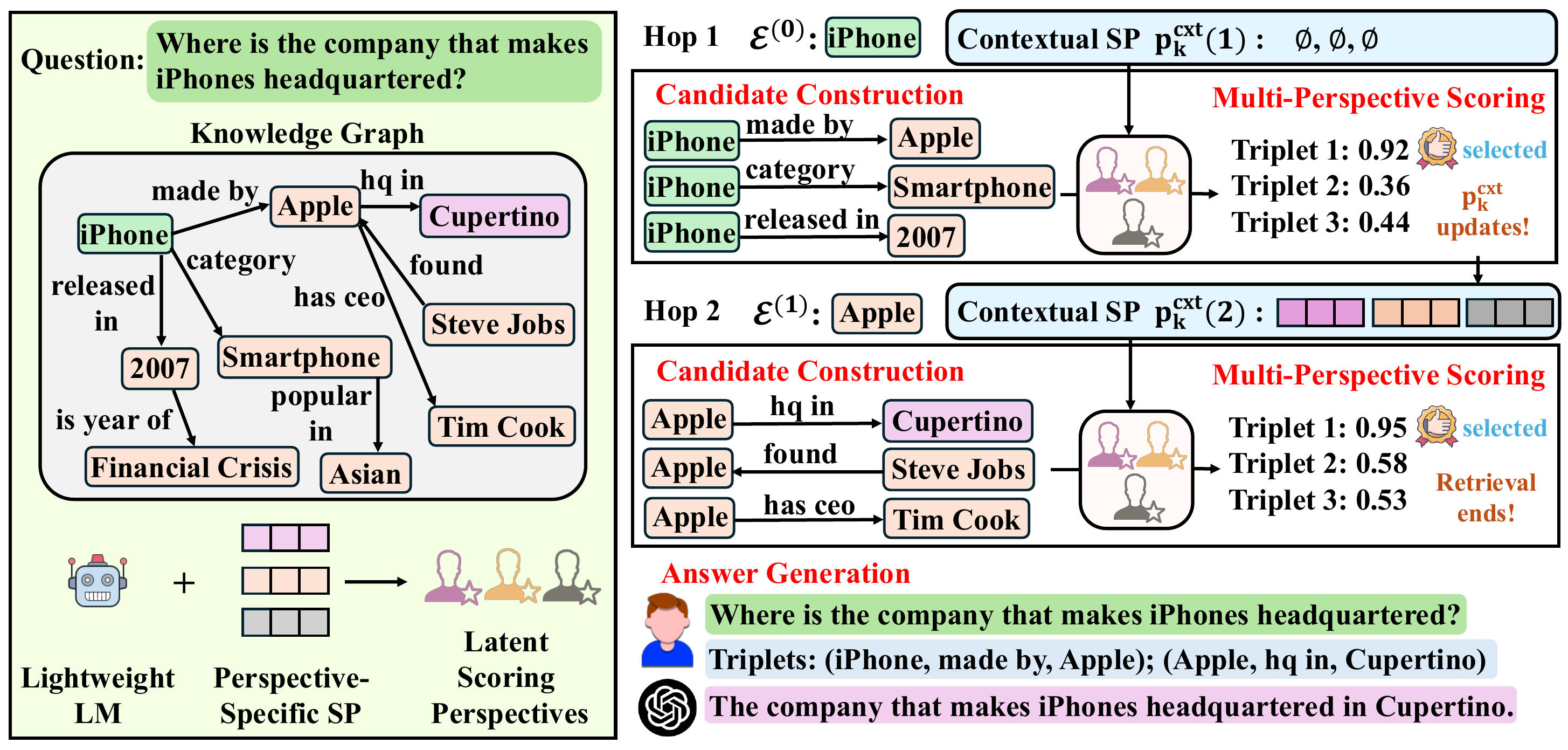}

    \caption{Overview of LSP-RAG framework for KG-RAG. The framework consists of: (1) Candidate Construction: From the frontier entities $\mathcal{E}^{(t-1)}$, initially the entities in question $q$, the framework iteratively expands the candidate triplet set $\mathcal{C}^{(t)}$ from the KG. (2) Multi-Perspective Prompted Scoring: Candidate triplets are evaluated by a lightweight language model (LM) with two types of soft prompts (SP): Perspective-specific soft prompts induce multiple latent scoring perspectives over candidate triplets. Contextual soft prompt is updated across hops to propagate reasoning history and maintain coherence during multi-hop inference. Perspective scores are adaptively aggregated into a final score for each candidate. Top-scoring triplets are selected at each hop and accumulated into the evidence set. (3) Answer Generation: The collected evidence is provided to a large generator for answer generation.}
    \vspace{-10pt}
  \label{fig:framework}
\end{figure*}

We propose LSP-RAG, a compact and adaptive framework for Knowledge Graph Retrieval-Augmented Generation (KG-RAG), designed to enhance multi-hop reasoning and knowledge retrieval using lightweight language models. LSP-RAG operates in a multi-hop retrieval setting, where relevant triplets from the knowledge graph are selected iteratively based on both the input query $q$ and the reasoning context of explored knowledge. As shown in Figure~\ref{fig:framework}, the framework introduces two key components: (1) a \textit{multi-perspective prompted scoring} module, where multiple latent scoring perspectives guided by distinct \emph{perspective-specific soft prompts} provide complementary prompt-conditioned scoring functions; and (2) \emph{contextual soft prompts} that are dynamically constructed from the query and previously explored knowledge, enabling hop-aware retrieval and maintaining coherence across multi-hop reasoning steps. These components work together to provide a flexible and context-sensitive evaluation of candidate triplets, allowing LSP-RAG to robustly retrieve useful knowledge without relying on enormous-scale LLMs.

\subsection{Multi-Hop Retrieval Framework}

Given a knowledge graph $\mathcal{G}$ and a natural language question $q$, our objective is to retrieve a set of triplets $\mathcal{T}_q^*$ from $\mathcal{G}$ that provide faithful evidence for answering $q$. To achieve this, we design an iterative retrieval framework that incrementally explores the KG in a step-wise manner.

\noindent\textbf{Step 1: Candidate Construction.}  
To initiate retrieval, we first identify a set of query entities $\mathcal{E}_q$ mentioned in the question $q$, and initialize the first-hop frontier as $\mathcal{E}^{(0)} = \mathcal{E}_q$.  
At the $t$-th retrieval hop, we construct a set of candidate triplets $\mathcal{C}^{(t)}$ by expanding outward from the current frontier $\mathcal{E}^{(t-1)}$.  
Specifically, we include all outgoing triplets $(e, r, e')$ for $e \in \mathcal{E}^{(t-1)}$, as well as reversed triplets $(e, r^{-1}, e')$ whenever $(e', r, e)$ appears in the KG:
\begin{equation}
\begin{aligned}
&\mathcal{C}^{(t)} = \{ (e, r, e') \mid e \in \mathcal{E}^{(t-1)} \} \;\cup\;\\
&\{ (e, r^{-1}, e') \mid (e', r, e) \in \mathcal{G}, \; e \in \mathcal{E}^{(t-1)} \}.
\end{aligned}
\end{equation}
This candidate construction step ensures broad coverage of possible evidence paths, supporting both forward and backward reasoning chains.  
The resulting set $\mathcal{C}^{(t)}$ serves as the input to the scoring module of hop $t$ described in the next step.

\noindent\textbf{Step 2: Multi-Perspective Scoring and Triplet Selection.}  
Each candidate triplet $x \in \mathcal{C}^{(t)}$ is evaluated for its relevance to the question $q$ and the context knowledge explored before.
To model heterogeneous relevance signals and maintain contextual consistency across hops, we employ a modular scoring mechanism based on multi-perspective prompted scoring with soft prompts (detailed in subsection~\ref{Multi-Perspective Scoring}).  
This module outputs a contextualized score $S(q,x,t)$ for each candidate.  
We retain the top-$M$ triplets by $S(q,x,t)$ for further exploration:
\begin{equation}
\mathcal{T}^{(t)}_{\text{sel}} = \text{Top-}M \{ S(q,x,t) \mid x \in \mathcal{C}^{(t)} \}.
\end{equation}
This step prioritizes the relevant triplets while controlling the expansion size and noise accumulation. 
We maintain a visited set $\mathcal{V}^{(t)}$ to track explored entities. At the beginning, we initialize $\mathcal{T}_q^* = \emptyset$ and $\mathcal{V}^{(0)} = \mathcal{E}^{(0)}$. At each hop $t$, the selected triplets $\mathcal{T}^{(t)}_{\text{sel}}$ are added to the evidence set, and the set of new tail entities is set as the frontier for the next hop:
\begin{align}
\mathcal{T}_q^* &\;\leftarrow\; \mathcal{T}_q^* \cup \mathcal{T}^{(t)}_{\text{sel}}, \\
\mathcal{E}^{(t)} &\;=\; \{ e' \mid (h, r, e') \in \mathcal{T}^{(t)}_{\text{sel}} \} \setminus \mathcal{V}^{(t-1)}, \\
\mathcal{V}^{(t)} &\;=\; \mathcal{V}^{(t-1)} \cup \mathcal{E}^{(t)}.
\end{align}
By keeping track of visited nodes, we avoid redundancy and loops.

\noindent\textbf{Iterative Retrieval.}  
The above process is repeated for a fixed number of hops $H$, progressively building an evidence graph rooted in the query. The final triplet set $\mathcal{T}_q^*$ contains the most relevant facts and is passed to the downstream answer generator.

Unlike methods that rely on large-scale language models to implicitly model reasoning chains, our framework performs retrieval in an explicit and structured way using only lightweight language models for triplet scoring.  
This enables efficient and scalable multi-hop retrieval with better evidence-level interpretability and lower cost.  
The design of our scoring module is key to enabling such structured reasoning and will be detailed in the next subsection.

\subsection{Multi-Perspective Prompted Scoring}
\label{Multi-Perspective Scoring}

To robustly evaluate candidate triplets in multi-hop retrieval, we design a compact neural scoring module based on multi-perspective prompted scoring with contextualized soft prompts. A \emph{soft prompt} refers to a learnable continuous embedding prepended to the input of a language model, allowing efficient task adaptation without fine-tuning the model parameters. This module addresses two key challenges in KG-RAG by applying different soft prompts: (1) modeling heterogeneous relevance signals with multiple latent scoring perspectives, and (2) maintaining contextual coherence across hops. To this end, each latent scoring perspective is equipped with two types of trainable soft prompts: a \emph{perspective-specific soft prompt} that parameterizes a prompt-conditioned scoring function, and a \emph{contextual soft prompt} that evolves over retrieval steps by summarizing the perspective's prior output representations. These components allow the model to flexibly score candidates while remaining sensitive to context evolution across hops.

Each latent scoring perspective operates as an independent scorer built on a shared, lightweight language model. Given a question $q$ and a candidate triplet $x$ at hop $t$, the $k$-th of $K$ latent scoring perspectives constructs its input by concatenating the \emph{perspective-specific soft prompt} $p_k^{\text{per}}$, the \emph{contextual soft prompt} $p_k^{\text{cxt}}(t)$, the question $q$, and a natural language serialization of the candidate triplet $x$ to obtain a new output embedding:
\begin{equation}
\mathbf{z}_k(q,x,t) = \mathrm{LM}\big([ p_k^{\text{cxt}}(t);\;p_k^{\text{per}};\; q;\; x]\big),
\end{equation}
where $\mathrm{LM}$ denotes the lightweight language model shared across all latent scoring perspectives. $p_k^{\text{per}}$ is a learnable embedding sequence that is unique to perspective $k$ and $p_k^{\text{cxt}}(t)$ encodes knowledge from the last hop. $\mathbf{z}_k(q,x,t)$ encodes prompt-conditioned relevance under the current perspective and reasoning state. 

To maintain temporal continuity, the \emph{contextual soft prompt} is updated after each hop by aggregating perspective outputs from the previous step:
\begin{equation}
p_k^{\text{cxt}}(t) = g\left( \frac{1}{|\mathcal{C}^{(t-1)}|} \sum_{x \in \mathcal{C}^{(t-1)}} \mathbf{z}_k(q,x,t{-}1) \right),
\end{equation}
where $g(\cdot)$ is a learned projection MLP layer. For the first hop ($t=1$), this prompt is omitted in the input of $\mathrm{LM}$. This update allows each latent scoring perspective to track its own contextual state across multiple retrieval steps. The $k$-th perspective transforms its embedding into a scalar score $s_k(q,x,t)\in(0,1)$ with a learnable linear layer $\mathbf{w}_k$ and a sigmoid layer $\sigma$:
\begin{equation}
s_k(q,x,t) = \sigma( \mathbf{w}_k^\top \mathbf{z}_k(q,x,t)),
\end{equation}
and an adaptive aggregation network assigns weights to each perspective. Specifically, the lightweight language model encoder produces a representation of the question and candidate as $f(q,x)$, and a trainable aggregation matrix $W_r$ maps this to a weight distribution on $K$ perspectives for triplet $x$:
\begin{equation}
\alpha(q,x,t) = \mathrm{softmax}(W_r \cdot f(q,x)).
\end{equation}
The final score for candidate $x$ is the weighted sum over latent scoring perspectives:
\begin{equation}
S(q,x,t) = \sum_{k=1}^{K} \alpha(q,x,t)[k] \cdot s_k(q,x,t).
\end{equation}

This scoring architecture directly addresses the key challenges in multi-hop KG retrieval. The \emph{perspective-specific soft prompts} induce multiple prompt-conditioned scoring functions, enabling flexible matching of heterogeneous relevance signals without requiring predefined semantic roles. Since supervision is provided at the level of triplet relevance rather than perspective labels, these latent scoring perspectives should not be interpreted as human-defined reasoning categories. Instead, they serve as latent scoring components that are adaptively aggregated for each question-triplet pair. The \emph{contextual soft prompts} preserve reasoning continuity across hops by incorporating step-wise retrieval history into scoring.  
Together, these components allow the retriever to adapt its behavior based on both the question semantics and the evolving context.  
Importantly, the entire module operates over lightweight language models with soft prompt tuning, offering an efficient and scalable alternative to LLM-based retrievers. 
The overall retrieval procedure is summarized in Algorithm~\ref{alg:lsprag}.

\begin{algorithm}[t]
\caption{LSP-RAG Retrieval (per question $q$; inference-time scoring with learned prompts)}
\label{alg:lsprag}
\begin{algorithmic}[1]
\REQUIRE KG $\mathcal{G}$, query $q$, hops $H$, top-$M$, latent scoring perspectives $K$
\REQUIRE Frozen LM encoder; learned parameters $\Theta=\{\,\{p_k^{\text{per}}\}_{k=1}^K,\;g,\;\{\mathbf{w}_k\}_{k=1}^K,\;W_r\,\}$
\STATE // \textit{Notes:} $p_k^{\text{per}}\in\mathbb{R}^{L_{\text{per}}\times d}$ (e.g., $L_{\text{per}}{=}5$ tokens), $p_{\text{init}}^{\text{cxt}}\in\mathbb{R}^{L_{\text{cxt}}\times d}$ (e.g., $L_{\text{cxt}}{=}1$ token), $f$ is the encoding of the LM encoder.
\STATE $\mathcal{E}^{(0)} \leftarrow \text{LinkEntities}(q)$,\quad $\mathcal{V}^{(0)} \leftarrow \mathcal{E}^{(0)}$,\quad $\mathcal{T}_q^* \leftarrow \emptyset$, \quad$p_{\text{init}}^{\text{cxt}}\leftarrow\emptyset$
\FOR{$k=1$ to $K$}
  \STATE $p_k^{\text{cxt}}(1) \leftarrow p_{\text{init}}^{\text{cxt}}$
\ENDFOR
\FOR{$t=1$ to $H$}
  \STATE $\mathcal{C}^{(t)} \leftarrow \text{ExpandCandidates}(\mathcal{G}, \mathcal{E}^{(t-1)})$
  \FORALL{$x \in \mathcal{C}^{(t)}$}
     \FOR{$k=1$ to $K$}
        \STATE $\mathbf{z}_k(q,x,t) \leftarrow \mathrm{LM}\!\big([\,p_k^{\text{cxt}}(t);\;p_k^{\text{per}};\;q;\;x\,]\big)$
        \STATE $s_k(q,x,t) \leftarrow \sigma(\mathbf{w}_k^\top \mathbf{z}_k(q,x,t))$
     \ENDFOR
     \STATE $\alpha(q,x,t) \leftarrow \mathrm{softmax}\!\big(W_r\, f(q,x)\big)$
     \STATE $S(q,x,t) \leftarrow \sum_{k=1}^{K} \alpha(q,x,t)[k] \cdot s_k(q,x,t)$
  \ENDFOR
  \STATE $\mathcal{T}^{(t)}_{\text{sel}} \leftarrow \text{Top-}M\ \text{candidates in }\mathcal{C}^{(t)}\ \text{by } S(q,\cdot,t)$
  \STATE $\mathcal{T}_q^* \leftarrow \mathcal{T}_q^* \cup \mathcal{T}^{(t)}_{\text{sel}}$
  \STATE $\mathcal{E}^{(t)} \leftarrow \{e' | (h,r,e') \in \mathcal{T}^{(t)}_{\text{sel}}\} \setminus \mathcal{V}^{(t-1)}$
  \STATE $\mathcal{V}^{(t)} \leftarrow \mathcal{V}^{(t-1)} \cup \mathcal{E}^{(t)}$
  \IF{$t < H$}
    \FOR{$k=1$ to $K$}
       \STATE $\displaystyle \bar{\mathbf{z}}_k(t) \leftarrow \frac{1}{|\mathcal{C}^{(t)}|}\sum_{x\in \mathcal{C}^{(t)}} \mathbf{z}_k(q,x,t)$
       \STATE $p_k^{\text{cxt}}(t{+}1) \leftarrow g\!\big(\bar{\mathbf{z}}_k(t)\big)$
    \ENDFOR
  \ENDIF
\ENDFOR
\ENSURE $\mathcal{T}_q^*$
\end{algorithmic}
\end{algorithm}

\subsection{Answer Generation}

After the multi-hop selection process, we obtain the final evidence triplet set $\mathcal{T}^*_q$ that aggregates promising triplets across hops. 
To generate the final answer, we leverage a powerful large-scale language model, such as ChatGPT, which excels at reasoning over textualized knowledge. Each triplet $x$ is first serialized into a natural language sentence. The evidence set is then concatenated into a textual context.
The LLM receives the question $q$ and the retrieved evidence $\mathcal{T}_q^*$ as input:
\begin{equation}
a_q = \mathrm{LLM}\big(\,[q ; \mathcal{T}_q^*]\,\big),
\end{equation}
where $a_q$ denotes the generated answer. 
Since the LLM is conditioned on the selected evidence, it grounds its reasoning on relevant multi-hop chains rather than the entire KG. 

While this generation step can leverage powerful LLMs, our framework is designed to minimize reliance on them. Due to the LSP-RAG retriever's multi-perspective prompted scoring mechanism built from compact language models and soft prompts, the system can already filter highly relevant and coherent evidence chains. This allows downstream generation to focus on language fluency, rather than compensating for missing knowledge or incorrect reasoning paths.

By explicitly addressing key challenges in KG-RAG, such as heterogeneous relevance estimation via latent scoring perspectives, cross-hop consistency via contextual soft prompts, and LLM scale dependency via lightweight model retrieval, our approach provides a scalable and interpretable solution. The triplet evidence offers transparent reasoning traces, while the latent scoring perspectives remain internal scoring components and the generation stage remains modular and adaptable to different deployment needs.

\subsection{Training Objective}

Our learnable components in LSP-RAG are trained by a contrastive learning process. Importantly, the underlying pretrained language models (both the lightweight language model and the large answer generator) remain frozen throughout training. This ensures that learning focuses on adapting the lightweight modules for scoring and information summarization.

Supervision is provided by question-answer pairs together with ground-truth subgraphs. For a given question $q$, in the hop $t$, let $\mathcal{T}^{+}_{q,t}\in \mathcal{C}^{(t)}$ denote the set of gold triplets that lie on reasoning paths leading a query entity in $\mathcal{E}_q$ to any answer entity within a predefined path length, and let $\mathcal{T}^{-}_{q,t}\in \mathcal{C}^{(t)}$ denote triplets sampled from the candidates that are not part of any gold path. These triplets introduce useful contextual information directly or indirectly relevant to the final answer. The objective is to assign higher scores to $\mathcal{T}^{+}_{q,t}$ while suppressing $\mathcal{T}^{-}_{q,t}$.

We adopt a cross-entropy loss over triplets at hop $t$:
\begin{equation}
\begin{aligned}
    \mathcal{L}& = - \frac{1}{|\mathcal{T}^+_{q,t}|}\sum_{x \in \mathcal{T}^+_{q,t}}  \log (S(q,x,t))
    \\
    &- \frac{1}{|\mathcal{T}^-_{q,t}|}\sum_{x \in \mathcal{T}^-_{q,t}}  \log (1-S(q,x,t)),
\end{aligned}
\end{equation}
All trainable parameters are optimized jointly by backpropagation.
Through this objective, the scorer learns to emphasize relevant triplets and suppress distractors, providing reliable multi-hop evidence for the downstream answer generator.

\section{Experiments}

We evaluate the effectiveness of our proposed approach across multiple QA benchmarks.

\subsection{Datasets}

We evaluate our method on four Question Answering datasets: WebQuestionsSP (WebQSP)~\cite{yih-etal-2016-value}, WikiMovies~\cite{miller-etal-2016-key}, and the datasets of MetaQA~\cite{metaqa} in both 1-hop and 2-hop settings. WebQSP contains 4,737 natural language questions annotated with corresponding SPARQL queries over Freebase~\cite{10.1145/1376616.1376746}, a broad-coverage knowledge graph. This dataset requires both single- and multi-hop structured reasoning and is commonly used to evaluate complex KG-based QA. WikiMovies is a domain-specific dataset focused on the movie domain, constructed from WikiData and Wikipedia, with around 100k question-answer pairs over 75,000 movie-related entities targeting factual reasoning over a curated movie KG. MetaQA builds upon WikiMovies by introducing paraphrased and more compositional questions. In the 1-hop setting, each question is grounded on a single KG fact, while the 2-hop version requires 2-hop inference to connect two supporting facts. Compared to WikiMovies, MetaQA includes more natural language variation and higher reasoning complexity, making it suitable for evaluating generalization under compositionality. Together, these datasets span both broad-domain and specialized-domain KGs, and cover reasoning from simple fact retrieval to multi-hop compositional queries. We follow the standard train/validation/test splits provided by each benchmark.

\subsection{Baselines}\label{sec:baseline} We evaluate our method against a broad set of baselines. These baselines differ in how they retrieve and reason over KG evidence, ranging from retrieval-free prompting to iterative graph reasoning with large LLMs. In contrast, our method achieves effective multi-hop reasoning using lightweight models. We evaluate our method against several widely-used and representative baselines. \textbf{IO-prompt}~\cite{gpt} uses ChatGPT-3.5 to directly generate answers without the assistance of any external evidence, often resulting in hallucinations or factual errors due to limited LLM internal knowledge. \textbf{CoT-prompt}~\cite{wei2022chain} improves upon this by encouraging the model to generate intermediate reasoning steps, but its performance is still bounded by the LLM's pretraining knowledge. \textbf{Self-Consistency}~\cite{wang2022self} further extends CoT by sampling multiple reasoning paths and selecting the most frequent answer, improving robustness but still lacking access to external facts. \textbf{Sim-Retrieve}~\cite{select} retrieves KG triplets by computing embedding similarity between questions and candidate triplets using LLaMA or Flan-T5. While simple and efficient, it often retrieves noisy or irrelevant facts, especially under multi-hop queries. \textbf{G-Retriever}~\cite{G-Retriever} formulates graph retrieval as a prize objective to select a compact, relevant subgraph as context, improving scalability and reducing hallucinations. \textbf{GNN-RAG}~\cite{GNN-RAG} uses a GNN as a dense subgraph reasoner to retrieve answer candidates, extracts KG paths connecting question entities to candidates, verbalizes these paths, and feeds them to an LLM for final reasoning. This improves faithfulness without extra LLM calls for retrieval. \textbf{GNN-Scoring} introduces a graph neural network (GNN) with the same size as GNN-RAG to score and filter triplets with a lightweight language model encoding the query within the input of the GNN, but it still struggles with long-range dependencies and lacks semantic adaptability. \textbf{Think-on-Graph (ToG)}~\cite{sun2024think} improves retrieval reasoning by iteratively selecting and scoring subgraphs using LLaMA, Flan-T5, or ChatGPT-3.5, and generating answers with ChatGPT-3.5. While effective, it incurs a higher computational cost and still depends heavily on the size of LLMs. \textbf{Reasoning on Graphs (RoG)}~\cite{rog} adopts a planning-retrieval-reasoning pipeline: it first generates KG-grounded relation-path plans, then retrieves instantiated reasoning paths from the KG to support faithful and interpretable reasoning. In contrast, our method uses lightweight models to score and retrieve, achieving accurate reasoning without relying on expensive models.

\begin{table*}[t]
\centering
\small
\caption{Experimental results (accuracy in \%) with standard deviation. We group methods into retrieval-free prompting, ChatGPT retrieval, and lightweight LM retrieval. The best result within each lightweight LM group is shown in \textbf{bold}. The best result among all compared methods is shown with an \underline{underline}.}

\label{tab:main_results}
\renewcommand{\arraystretch}{1.6}
\setlength\tabcolsep{8.0pt}
\begin{tabular}{llcccc}
\hline
\textbf{Category} & \textbf{Method} & \textbf{WebQSP} & \textbf{WikiMovies} & \textbf{MetaQA-1hop} & \textbf{MetaQA-2hop} \\
\hline

\multirow{3}{*}{Retrieval-free} 
  & IO-prompt & 62.32 $\pm$ 0.41 & 69.24 $\pm$ 0.35 & 60.02 $\pm$ 0.60 & 23.01 $\pm$ 0.79 \\
  & CoT-prompt & 62.74 $\pm$ 0.46 & 72.45 $\pm$ 0.42 & 64.98 $\pm$ 0.52 & 26.71 $\pm$ 0.88 \\
  & Self-Consistency & 61.11 $\pm$ 0.53 & 73.64 $\pm$ 0.37 & 71.38 $\pm$ 0.39 & 30.52 $\pm$ 1.01 \\
\hline

\multirow{2}{*}{ChatGPT retrieval } 
  & ToG  & 74.24 $\pm$ 0.91 & \underline{95.77} $\pm$ 0.63 & \underline{96.33} $\pm$ 0.77 & \underline{94.43} $\pm$ 0.89 \\
   & RoG  & \underline{74.27} $\pm$ 0.94 & 95.62 $\pm$ 0.51 & 96.15 $\pm$ 0.60 & 94.06 $\pm$ 0.92 \\
\hline

\multirow{7}{*}{LLaMA2 retrieval} 
  & Sim-Retrieve & 46.98 $\pm$ 1.89 & 50.09 $\pm$ 1.03 & 47.44 $\pm$ 0.99 & 21.02 $\pm$ 1.93 \\
  & G-retriever
 & 69.02
 $\pm$ 1.44 & 80.95
 $\pm$ 1.11 & 77.77
 $\pm$ 0.92 & 67.08
 $\pm$ 1.67 \\
  & GNN-Scoring & 63.86 $\pm$ 0.73 & 78.32 $\pm$ 0.48 & 83.82 $\pm$ 0.52 & 71.63 $\pm$ 0.91 \\
  & GNN-RAG
 & 63.43
 $\pm$ 0.34 & 77.52
 $\pm$ 0.62 & 82.89
 $\pm$ 0.63 & 70.35
 $\pm$ 0.84 \\
  & ToG & 66.70 $\pm$ 0.85 & 86.62 $\pm$ 0.59 & 88.63 $\pm$ 0.67 & 78.01 $\pm$ 0.93 \\
  & RoG & 64.19
 $\pm$ 0.89 & 85.14
 $\pm$ 0.74 & 86.08
 $\pm$ 0.73 & 76.34
 $\pm$ 0.90 \\
  & \cellcolor{gray!21}{\textbf{LSP-RAG}} & \cellcolor{gray!21}{\textbf{75.02 $\pm$ 0.88}} & \cellcolor{gray!21}{\textbf{98.29 $\pm$ 0.76}} & \cellcolor{gray!21}{\textbf{97.25 $\pm$ 0.99}} & \cellcolor{gray!21}{\textbf{94.91 $\pm$ 0.72}} \\
\hline
\multirow{7}{*}{LLaMA3 retrieval} 
  & Sim-Retrieve & 47.16
 $\pm$ 1.01 & 51.02
 $\pm$ 1.08 & 47.46
 $\pm$ 0.70 & 22.96
 $\pm$ 1.49 \\
  & G-retriever
 & 70.12
 $\pm$ 1.74 & 82.04
 $\pm$ 1.32 & 76.21
 $\pm$ 0.90 & 69.23
 $\pm$ 1.87 \\
  & GNN-Scoring & 65.77
 $\pm$ 0.71 & 80.24
 $\pm$ 0.50 & 86.12
 $\pm$ 0.47 & 74.92
 $\pm$ 0.99 \\
  & GNN-RAG
 & 67.49
 $\pm$ 0.75 & 80.25
 $\pm$ 0.66 & 86.01
 $\pm$ 0.59 & 73.82
 $\pm$ 0.84 \\
  & ToG & 64.60
 $\pm$ 0.90 & 88.44
 $\pm$ 0.77 & 90.02
 $\pm$ 0.81 & 83.89 $\pm$ 0.96 \\
  & RoG & 66.99
 $\pm$ 0.94 & 87.93
 $\pm$ 0.72 & 89.79
 $\pm$ 0.81 & 82.02
 $\pm$ 0.90 \\
  & \cellcolor{gray!21}{\textbf{LSP-RAG}} & \cellcolor{gray!21}{\textbf{76.63 $\pm$ 0.53}} & \cellcolor{gray!21}{\textbf{98.33 $\pm$ 0.62}} & \cellcolor{gray!21}{\textbf{97.65 $\pm$ 1.08}} & \cellcolor{gray!21}{\textbf{94.99 $\pm$ 0.78}} \\
\hline
\multirow{7}{*}{Flan-T5 retrieval} 
  & Sim-Retrieve & 44.78 $\pm$ 1.35 & 52.39 $\pm$ 0.99 & 47.93 $\pm$ 0.85 & 20.17 $\pm$ 1.72 \\
  & G-retriever
 & 66.03 $\pm$ 1.51 & 79.11 $\pm$ 1.04 & 76.21 $\pm$ 0.88 & 69.23 $\pm$ 1.47 \\
  & GNN-Scoring & 64.56 $\pm$ 0.65 & 79.10 $\pm$ 0.44 & 84.69 $\pm$ 0.49 & 73.68 $\pm$ 0.86 \\
  & GNN-RAG
 & 62.51 $\pm$ 0.62 & 78.06 $\pm$ 0.38 & 83.98 $\pm$ 0.54 & 73.86 $\pm$ 0.91 \\
  & ToG & 72.25 $\pm$ 0.77 & 94.80 $\pm$ 0.78 & 90.67 $\pm$ 0.66 & 89.60 $\pm$ 0.98 \\
  & RoG & 50.95 $\pm$ 1.57 & 96.11 $\pm$ 0.66 & 90.12 $\pm$ 0.68 & 88.18 $\pm$ 0.91 \\
  & \cellcolor{gray!21}{\textbf{LSP-RAG}} & \cellcolor{gray!21}{\textbf{88.59 $\pm$ 0.83}} & \cellcolor{gray!21}{\textbf{98.22 $\pm$ 0.74}} & \cellcolor{gray!21}{\textbf{96.84 $\pm$ 0.87}} & \cellcolor{gray!21}{\textbf{94.95 $\pm$ 0.95}} \\
\hline
\end{tabular}
\vspace{-5pt}
\end{table*}

\subsection{Experimental Setup}

To evaluate the effectiveness of our proposed framework, we conduct experiments using \textbf{ChatGPT-3.5}~\cite{chatgpt2022}, \textbf{LLaMA2-7B}, \textbf{LLaMA3-8B}~\cite{llama2}, and \textbf{Flan-T5-3B}~\cite{t5} as backbone language models for retrieval and answering. In our framework, LLaMA and Flan-T5 are used as retriever models for scoring and selecting KG triplets, while ChatGPT-3.5 is used only during the final answer generation stage for some baselines. 
For triplet selection, the multi-perspective prompted scoring module is configured with $K=3$ latent scoring perspectives by default. The multi-hop expansion process is controlled by a hop budget $H=1$ for MetaQA-1hop and $H=2$ for other datasets, balancing retrieval depth and noise. We cap the number of retrieved triplets to 20 per query to maintain prompt length constraints. Perspective-specific and contextual soft prompts are jointly used to condition perspective scoring and improve relevance estimation under multi-hop contexts. 
All experiments are conducted with 5 random runs, and we report the average accuracy to mitigate variance. 

\subsection{Backbone Models}

Our retrieval module operates on top of lightweight pre-trained language models, enabling efficient and scalable scoring without relying on large generative models during retrieval. We experiment with three representative retriever backbones: LLaMA2-7B, LLaMA3-8B, and Flan-T5-3B.

\noindent\textbf{LLaMA.}
LLaMA~\cite{llama2} is a family of decoder-only transformer models pre-trained on large-scale public corpora. We adopt the 7B/8B variants of LLaMA2/3 as feature extractors for triplet scoring. In our setup, the LLaMA parameters are frozen, and only the soft prompts, projection layers, and lightweight multi-perspective scoring components are trained. As a decoder-only model, LLaMA encodes the concatenated prompt and triplet input into hidden representations, from which we use the latent output representation for downstream scoring.

\noindent\textbf{Flan-T5.}
Flan-T5~\cite{t5} is an instruction-tuned extension of the T5 encoder-decoder architecture. The Flan variant is trained on a large collection of prompting tasks, making it effective under instruction-style input formats. We use the 3B variant in encoder-only mode to encode the concatenated soft prompts, question, and candidate triplet. Similar to LLaMA, the backbone parameters of Flan-T5 are frozen throughout training, with only the soft prompts, projection layers, and lightweight scoring components updated.

\noindent\textbf{Soft Prompt Integration.}
For both model families, we prepend soft prompts as continuous embedding vectors at the input layer. This design does not modify the tokenizer or the internal architecture of the backbone language model. During training, gradients are propagated through the perspective-specific soft prompts, contextual soft prompts, projection layer, and adaptive aggregation module. This makes LSP-RAG compatible with different pre-trained language models while keeping the backbone frozen.

\subsection{Training Details} We train all models using the Adam optimizer with learning rate $1 \times 10^{-4}$, batch size 20, and linear learning rate decay over 5 epochs. Prompt parameters are initialized randomly and trained from scratch. Each latent scoring perspective uses two separate prompt embeddings: one perspective-specific and one contextual. The underlying language models are kept frozen throughout training, and only the soft prompts, projection layers, and lightweight scoring components are updated. We train all experiments on 4 NVIDIA A100 GPUs with 80GB memory using mixed-precision training enabled via PyTorch. During inference, retrieval is performed up to $H=1$ hop for the MetaQA-1hop dataset and $H=2$ for others, with $M=20$ triplets selected per hop. We set the number of latent scoring perspectives to $K=3$. The dimension of the latent embeddings in our method is set to the same dimension as the retrieval LLM, i.e., 2,048 for Flan-T5-3B and 4,096 for LLaMA2/3. All KG triplets are preprocessed into a bidirectional index, which maps each entity to its outgoing and incoming relations. This index is loaded into memory as a dictionary for efficient candidate expansion at each hop. \subsection{Evaluation Protocol} We follow standard entity-level accuracy for all datasets, considering a prediction correct if it matches any of the gold answers. Each model is evaluated with 5 different random seeds, and we report the average accuracy and standard deviation across runs. For fair comparison, all baseline methods use the same KG and entity linker. The knowledge graph used for WikiMovies and WebQSP is provided by~\cite{sun2018open}, while for MetaQA we use the set from~\cite{kim-etal-2023-kg}. ToG baselines are re-implemented with the same number of hops and selection budget as LSP-RAG. 

Our implementation is based on PyTorch and Python 3.11.7. We release our code at \url{https://github.com/HaochenLiu2000/LSP-RAG}.

\subsection{Results and Analysis}

Table~\ref{tab:main_results} shows that LSP-RAG consistently achieves the highest accuracy across all datasets and model backbones, significantly outperforming both retrieval-free prompting methods and alternative retrievers. Retrieval-free baselines such as IO-prompt and CoT-prompt fall short due to their lack of access to structured evidence. While embedding similarity-based retrievers often fail to model multi-hop dependencies, GNN-based retrievers cannot capture semantic alignment. Despite modeling relationships of question and knowledge, they provide limited flexibility in matching complex, compositional semantics. In contrast, LSP-RAG demonstrates strong performance even under lightweight language models, outperforming ToG/RoG even when using ChatGPT as its retriever. This highlights the effectiveness of LSP-RAG: the perspective-specific soft prompts induce multiple latent scoring perspectives for flexible relevance estimation, while the hop-aware contextual soft prompts maintain coherence across reasoning steps. Interestingly, Flan-T5-3B achieves substantially higher accuracy on WebQSP, suggesting that the encoder architecture like Flan-T5 is particularly well-suited for latent triplet scoring and multi-hop relevance estimation in the open domain. Importantly, LSP-RAG requires no LLM finetuning or large-scale language model inference during retrieval, showing that compact models, when equipped with the right inductive biases, can also support accurate and robust reasoning.

\vspace{-8pt}
\subsection{Parameter Study}\label{sec:param}

\begin{figure*}[t]
    \centering
    \includegraphics[width=0.7\linewidth]{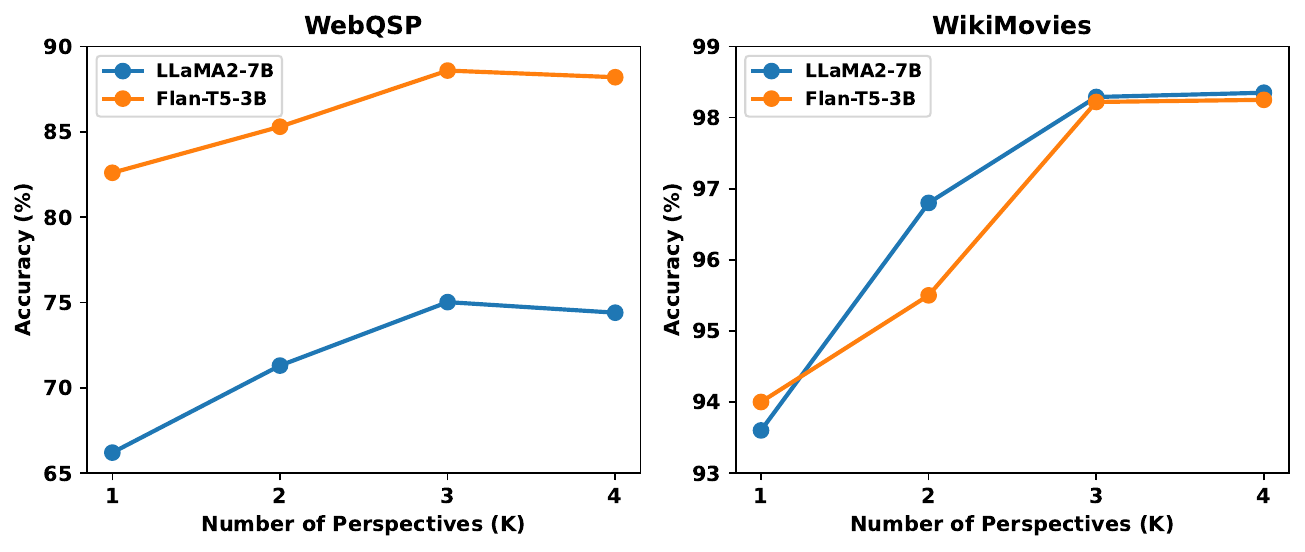}

    \caption{
    QA accuracy with varying numbers of latent scoring perspectives $K$.
    We report results on WebQSP and WikiMovies using both LLaMA2-7B and Flan-T5-3B. 
    Increasing $K$ from 1 to 3 improves performance across all settings, as multiple perspectives provide additional prompt-conditioned scoring capacity. 
    Performance saturates or slightly drops at $K=4$, indicating that a small number of latent scoring perspectives is sufficient for flexible triplet scoring without unnecessary redundancy.
    }

    \label{fig:para}
\end{figure*}

\begin{table*}[t]
\centering
\small
\caption{Ablation results (accuracy in \%) comparing the full LSP-RAG model with variants that remove the perspective-specific (per) and/or contextual (cxt) soft prompts. We report accuracy on four datasets: WebQSP (WQ), WikiMovies (WM), MetaQA-1hop (MQ-1), and MetaQA-2hop (MQ-2). The best results are shown in \textbf{bold}.}
\label{tab:ablation}
\renewcommand{\arraystretch}{1.6}
\setlength\tabcolsep{8.0pt}
\begin{tabular}{l|cccc|cccc|cccc}
%\toprule
\hline
\multirow{2}{*}{\textbf{Method}} 
& \multicolumn{4}{c|}{\textbf{LLaMA2-7B}} 
& \multicolumn{4}{c|}{\textbf{LLaMA3-8B}}
& \multicolumn{4}{c}{\textbf{Flan-T5-3B}} \\
%\cmidrule(lr){2-9}% \cmidrule(lr){6-9}
& WQ & WM & MQ-1 & MQ-2
& WQ & WM & MQ-1 & MQ-2 
& WQ & WM & MQ-1 & MQ-2\\
\hline
\cellcolor{gray!21}{Full Model}            
& \cellcolor{gray!21}{\textbf{75.02}} & \cellcolor{gray!21}{\textbf{98.29}} & \cellcolor{gray!21}{\textbf{97.25}} & \cellcolor{gray!21}{\textbf{94.91}} 
& \cellcolor{gray!21}{\textbf{76.63}} & \cellcolor{gray!21}{\textbf{98.33}} & \cellcolor{gray!21}{\textbf{97.65}} & \cellcolor{gray!21}{\textbf{94.99}} & \cellcolor{gray!21}{\textbf{88.59}} & \cellcolor{gray!21}{\textbf{98.22}} & \cellcolor{gray!21}{\textbf{96.84}} & \cellcolor{gray!21}{\textbf{94.95}} \\
w/o per  
& 67.83 & 94.27 & 88.14 & 84.63 
& 68.62 & 95.01 & 92.13 & 90.00
& 83.74 & 95.08 & 90.82 & 88.94 
\\
w/o cxt     
& 70.01 & 95.46 & N/A    & 90.91 
& 71.04 & 95.47 & N/A    & 91.48
& 86.23 & 97.31 & N/A    & 91.18 \\
w/o Both              
& 63.98 & 86.01 & N/A & 77.86 
& 65.01 & 87.35 & N/A & 79.71
& 71.65 & 93.88 & N/A & 86.97 \\
%\bottomrule
\hline
\end{tabular}
\vspace{-15pt}
\end{table*}
\vspace{-5pt}

We further study the impact of the number of latent scoring perspectives $K$ in the multi-perspective prompted scoring module, focusing on WebQSP and WikiMovies with both LLaMA2-7B and Flan-T5-3B as the LLMs for retrieval.  
The results in Figure~\ref{fig:para} show a clear upward trend when increasing $K$ from 1 to 3: the models consistently achieve better QA accuracy as multiple latent perspectives provide additional prompt-conditioned scoring capacity. However, the benefit quickly saturates beyond this point. In fact, while $K=4$ performs slightly better than $K=3$ on WikiMovies, the improvement is marginal, and on WebQSP with LLaMA2-7B the accuracy at $K=4$ even drops compared to $K=3$.  

This observation suggests that a moderate number of latent scoring perspectives is sufficient to provide flexible relevance estimation for effective triplet scoring. Using too many perspectives may introduce redundancy and increases training as well as inference cost, without yielding consistent performance gains. For this reason, we set $K=3$ as the default configuration in our method, which achieves nearly the same accuracy as $K=4$ while being more cost-effective and requiring lower computing cost.

\subsection{Ablation Study}

We examine the impact of removing two key components: the perspective-specific soft prompts and the contextual soft prompts. Results in Table~\ref{tab:ablation} show that across all datasets and LLM settings, removing either consistently degrades performance. Without perspective-specific prompts, the model collapses multiple latent scoring perspectives into a less flexible scoring function, leading to weaker discrimination especially on multi-hop datasets such as WebQSP and MetaQA-2hop where heterogeneous relevance signals must be handled. Without contextual prompts, the model loses continuity across hops and struggles to select the right chain of evidence. This becomes severe in multi-hop reasoning QA settings. Notably, removing either type of prompt leads to consistent performance drops across both models. Removing perspective-specific soft prompts results in a larger degradation on both models, particularly on complex reasoning settings like WebQSP, indicating the importance of maintaining multiple prompt-conditioned scoring functions. In contrast, the removal of contextual soft prompts hurts both models on multi-hop datasets, highlighting the importance of preserving inter-hop coherence for complex reasoning. Both components play complementary and indispensable roles in enabling robust multi-hop retrieval.

\vspace{-5pt}
\section{Related Works}

\subsection{Retrieval-Augmented Generation}

Retrieval-Augmented Generation (RAG)~\cite{lewis2020retrieval} combines LLMs with external memory to address knowledge staleness and hallucination issues. The framework retrieves relevant passages from external corpora and conditions generation on both the input query and retrieved context. Early variants include end-to-end training approaches~\cite{izacard2023atlas}, dense passage retrieval optimization~\cite{karpukhin2020dense}, and fusion-in-decoder architectures~\cite{izacard2020leveraging}. Recent advances have addressed key limitations through sophisticated retrieval strategies: Self-RAG~\cite{asai2024self} enables adaptive retrieval decisions through reflection tokens, RAPTOR~\cite{sarthi2024raptor} constructs hierarchical tree structures via recursive clustering, Anthropic's Contextual Retrieval~\cite{anthropic2024contextual} reduces retrieval failures through contextual chunk embedding, GraphRAG~\cite{edge2024local} leverages knowledge graphs for global reasoning, and CRAG~\cite{yan2024corrective} introduces retrieval quality assessment mechanisms. Other developments include HyDE~\cite{gao2023precise} and agentic systems~\cite{singh2025agentic,dong2025rag}. However, most methods are text-centric and do not fully leverage structured knowledge graphs, limiting the reasoning transparency and multi-hop capability.

\vspace{-4pt}
\subsection{Conditional Computation and Multi-View Scoring}

Mixture of Experts (MoE) has emerged as a powerful architecture for scaling model capacity while maintaining computational efficiency through conditional computation~\cite{fedus2022switch,cai2024survey}. Originally proposed to divide problem spaces into homogeneous regions handled by specialized sub-networks~\cite{jacobs1991adaptive}, MoE has been successfully adapted to large language models where different experts are routed conditionally based on the input. Recent advances in MoE architectures include Mixtral 8x7B~\cite{jiang2024mixtral}, which demonstrated that sparse MoE models can outperform dense counterparts on knowledge-intensive tasks, and Switch Transformer~\cite{fedus2022switch}, which scales to trillion-parameter models by routing tokens to different expert networks through learned gating functions. However, MoE models have been shown to excel at memorization-intensive tasks but struggle with complex reasoning compared to dense models~\cite{muqeeth2024mixture}, suggesting that conditional computation alone does not necessarily provide interpretable reasoning ability. In knowledge graph reasoning specifically, MoMoK~\cite{zhang2024momok} introduces relation-guided modality knowledge experts for multi-modal knowledge graph completion to handle different relational contexts and modality-specific signals.

Different from conventional MoE architectures that rely on specialized expert networks, our work uses lightweight latent scoring perspectives induced by perspective-specific soft prompts. These perspectives are not designed as human-defined semantic experts; instead, they provide multiple prompt-conditioned scoring functions that are adaptively aggregated for KG triplet relevance estimation. This design preserves the efficiency benefits of lightweight conditional scoring while keeping the reasoning trace explicit through retrieved KG evidence paths, offering an effective way for KG-RAG.

\vspace{-4pt}
\section{Limitations}

Although LSP-RAG achieves strong retrieval and QA performance with lightweight language models, it still has several limitations. First, its computational cost increases with the number of latent scoring perspectives, since each candidate triplet is evaluated by multiple prompt-conditioned components. Second, LSP-RAG relies on hop-wise KG traversal with a fixed hop budget, which may miss evidence requiring long, indirect, or weakly connected reasoning paths. Third, our experiments mainly focus on relatively clean and well-structured KGs; extending the framework to incomplete, noisy, or large-scale heterogeneous graphs remains future work. Finally, latent scoring perspectives are not intended to correspond to human-defined semantic categories. The interpretability of LSP-RAG mainly comes from explicit KG triplets and hop-wise evidence paths, while controllable or semantically supervised scoring perspectives are left for future work.
\vspace{-4pt}
%\section{Ethical Considerations}

%LSP-RAG is designed to improve the factual grounding of LLM-based question answering by retrieving explicit evidence from knowledge graphs. However, the system may still produce incorrect or incomplete answers if the underlying KG contains outdated, biased, noisy, or missing facts. Since the final answer is generated by an LLM conditioned on retrieved triplets, hallucinations may still occur when the retrieved evidence is insufficient or ambiguous. Therefore, LSP-RAG should not be used as the sole decision-making component in high-stakes domains such as healthcare, law, finance, or public policy without human verification. In addition, the KGs and QA benchmarks used in this work may reflect biases from their original data sources. Future deployments should consider data provenance, privacy, fairness, and mechanisms for auditing retrieved evidence.

\section{Conclusion}
In this work, we proposed a novel knowledge graph-augmented retrieval framework for LLM-based question answering called LSP-RAG. Our approach introduces a multi-perspective prompted scoring module that leverages both perspective-specific and contextual soft prompts to evaluate candidate triplets, enabling robust multi-hop retrieval across knowledge graphs. Through the designed hop-aware retrieval and selection process, our method ensures contextual and adaptive evidence exploration while maintaining efficiency. Importantly, the design does not require fine-tuning large backbone language models; instead, it relies on a compact scoring architecture while still fully exploiting the capabilities of LLMs for final answer generation. Extensive experiments on the benchmarks demonstrate the effectiveness of our approach. It consistently outperforms retrieval-based baselines, including ToG/RoG, even when the latter are powered by larger models such as ChatGPT. Our ablation studies further confirm the critical roles of perspective-specific and contextual prompts in improving multi-hop retrieval performance.  
Overall, our retrieval and scoring mechanisms enhance the reasoning ability of LLMs without relying on large models. We believe that LSP-RAG provides a general and cost-effective paradigm for future research in knowledge-enhanced LLM reasoning.
\vspace{-3pt}
\section{Acknowledgement}
This work was supported in part by the National Science Foundation (NSF) under Grants IIS-2144209, IIS-2223769, BCS-2228534, CMMI-2411248, and IIBR-2601942; by the Office of Naval Research (ONR) under Grant N000142412636; and by the Commonwealth Cyber Initiative (CCI) under Grant HV-4Q26-073.

\bibliography{iclr2026_conference,guanghui}
\bibliographystyle{IEEEtran}

\end{document}